\documentclass[twocolumn]{article}
\usepackage{graphicx}
\usepackage{eqnarray}
\usepackage{amssymb}
\usepackage{anysize}
\usepackage{amsfonts,amssymb,latexsym,amsmath,amscd}
\marginsize{25mm}{20mm}{35mm}{40mm} \textheight=215mm

\begin{document}
    \title{\textbf{Analysis and Design of an Organic High Speed Digital Electro-Optic Switch}}
    \author{AliAkbar Wahedy Zarch$^{1}$\footnote{Material presented in this
    paper is a part of AliAkbar Wahedy Zarch work on his
    thesis towards Ph.D degree. Dr. Hassan Kaatuzian is his supervisor on thesis.
    fatemahali@aut.ac.ir} \and
    Hassan Kaatuzian$^{1}$\and Ahmad Ajdarzadeh Oskouei$^{2}$ \and Ahmad Amjadi$^{3}$
    \\ \\ \normalsize $^{1}$Photonics Research Laboratory (PRL), Electrical Engineering Department\\
    \normalsize AmirKabir University of Technology, Tehran, IRAN,
    hsnkato@aut.ac.ir.\\
    $^{2}$\normalsize Laboratory of Ultrafast Spectroscopy, Ecole Polytechnique Federale
    de Lausanne(EPFL),\\\normalsize Switzerland, ahmad.ajdarzadeh@epfl.ch\\
    $^{3}$\normalsize Physics Department, Sharif University of Technology, Tehran,
    IRAN, amjadi@sharif.edu.}
    \date{}

    \maketitle

    \begin{abstract}
       We have analyzed and designed  an organic high speed digital optical switch (DOS) based on transverse
       electro-optic effect. In analysis section, we proposed a
       quantum photonic model (QPM) to explain linear electro-optic (EO) effect. This model interpret this effect by
       photon-electron interaction in attosecond regime. We
       simulate applied electric field on molecule and crystal by
       Monte-Carlo method in time domain. We show how a waveguide response to an optical
       signal with different wavelengths when a transverse electric field applied to the waveguide.
       In design section, we configure conceptually a $2\times2$ EO
       switch with full adiabatic coupler. In this DOS, we use a
       rib waveguides that its core has been constructed from NPP
       crystal. This switch is smaller at least to one-half of similar DOS in
       dimensions.
    \end{abstract}

    \vspace{2pc} \noindent{\it Keywords}: Quantum-photonic model(QPM),
        $\pi$-electron system, photon-electron
        interaction, attosecond regime, electro-optic (EO) effect,
        digital optical switch (DOS), $2\times2$ organic EO switch, NPP crystal.\\

    \section{\large Introduction}
        High-speed and high-performance optical switches are vital for optical
        communication networks and optical signal-processing systems. Among these, optical switches based on the
        electro-optic(EO) effect are promising for applications that require
        high-speed low-loss switching such as: optical burst switching, optical packet
        switching or EO beam deflector switching \cite{1}-\cite{16}. Mach–Zehnder interferometer
        (MZI) switches and directional coupler(DC) switches have been under intensive investigation for a long
        time, but they both have very limited fabrication tolerance.
        Furthermore, bias control networks and precise switching voltages
        are needed in MZI switches and DC switches to achieve high
        performance \cite{1,14}. Electro-optic polymers are particularly interesting for new device
        design and high-speed operation \cite{1}-\cite{7}. Organic optical materials like MNA, NPP, MAP have a high figure
        of merit in optical properties in comparison with inorganic optical materials such as BBO, LiNbO$_{3}$, \cite{17}.
        2-methyl-4-nitroaniline (MNA) and N-(4-nitrophenyl)-L-prolinol (NPP) have the highest figure
        of merit between organic nonlinear optical materials \cite{18}. Thus they are used for electro-optic
        and nonlinear optic applications\cite{17}-\cite{31}.
        Because of the large electro-optic coefficient of organic material, a certain amount of
        refractive index (RI) change can be realized with lower driving voltage
        than in other EO materials($1.5 V/\mu m$ for NPP,\cite{19}). Since there is better optical/RF
        velocity matching inside the EO organic, high-speed switching can
        be achieved easily.
        Digital optical switches (DOSs) are based on adiabatic
        propagation or modal evolution effect\cite{8}. These switches are
        wavelength-polarization insensitivity and electronics for control
        of these devices are simple. These switches can be applicable in optical network of phase array radar systems.
        Their structure of waveguides can be chosen
        the buried-type waveguide or
        rib-type (ridge-type) waveguide.
        In this paper we analyze and design an organic DOS that is based on EO phenomenon.
        Our favorite organic molecule is
        NPP. In analysis, we use QPM, to explain linear optical
        phenomena in molecular scales. We show that the phase retardation
        of input light with different wavelengths is distinctive, when it travels through the waveguide.
        In design we design a $2\times2$ DOS with adiabatic couplers. The main advantages of this
        switch are wide optical bandwidth, low power consumption
        and high speed data transmission.

    \section{\large Analysis}
        Some approaches are used to analyze and design
        optical switches in micrometer and nanometer sizes by quantum mechanics (QM) \cite{32}.
        In this paper an EO switch
        has been analyzed using QPM. This model gives us a constitutive
        vision about phenomena and real materials \cite{33}-\cite{36}.
        This approach is based on four elements: 1-Quasi-classic principle for justification of optical phenomenon in
        molecular scales, 2- Knowledge of crystal network and its space shape, 3-
        Short range intramolecular and intermolecular forces, 4-
        Monte-Carlo time domain simulation. In this model we suppose a laser beam is a
        flow of photons while passing through single crystal film, interacts
        with delocalization $\pi$-electron system of organic molecule and delays
        the photon in every layer \cite{36}. By precise calculation of these
        retardation in every layer, we obtain RI in
        specific wavelength and explain EO effect too. After one, we apply an
        external electric field to crystal to change its RI
        a little to switch specific wavelength to identified
        channel.

    \subsection{\normalsize Crystal and Molecular Structure of our Favorite Organic Compound}
\begin{figure}
            \centering
            \includegraphics[scale=0.22]{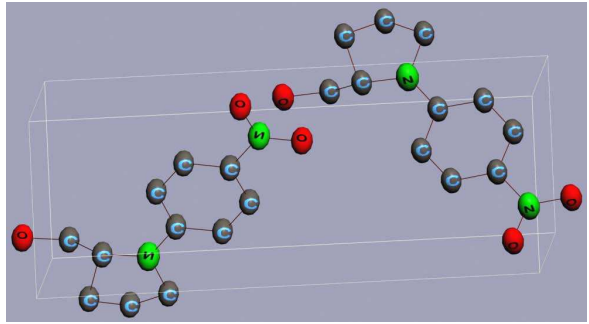}
            \caption{\footnotesize The crystal packing of NPP. This molecule has a monoclinic crystal
            structure with space group $P2_{1}$. There are two molecules in every cell.}\label{Fig2}
        \end{figure}

\begin{figure*}
            \centering
            \includegraphics[scale=1.2]{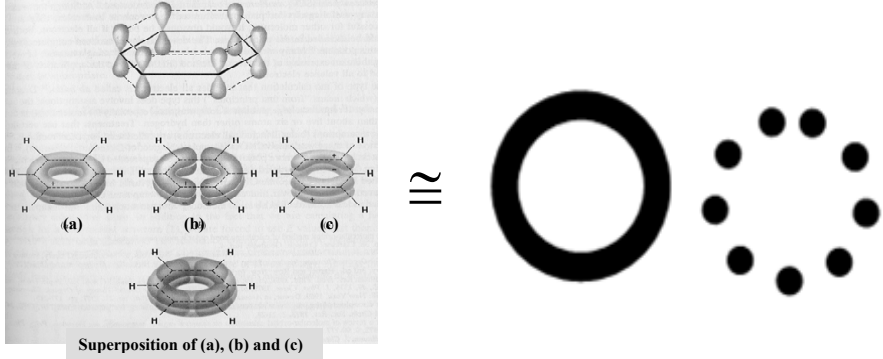}
            \caption{\footnotesize Electron cloud for Benzene molecule that obtained from Huckle theory \cite{37} and its
            approximation.}\label{Fig3}
        \end{figure*}

\begin{figure}
            \centering
            \includegraphics[scale=0.35]{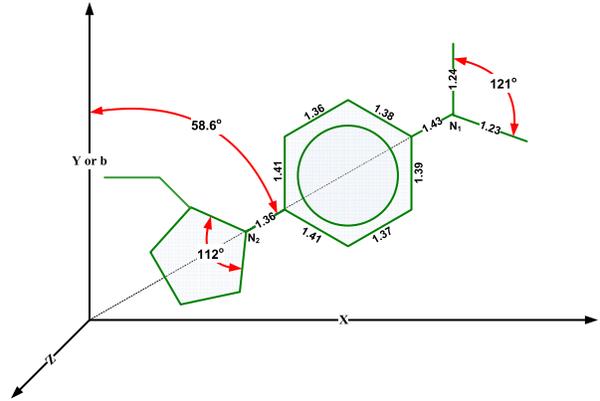}
            \caption{\footnotesize The NPP molecule in dielectric frame.
            The bond lengths (in angstrom) and angles of NPP molecule have been shown
            too. As we see the angle between charge transfer axis and Y axis
            is $58.6^{o}$. The mean plane of molecule is in X-Y plane approximately
            ($11^{o}$ accurately).}\label{Fig4}
        \end{figure}

        NPP $(C_{11}H_{14}N_{2}O_{3})$ crystallizes in the solid
        state in an acentric monoclinic (with space group $P2_{1}$) structure and
        their parameters are: a=5.261$A^{\circ}$,b=14.908$A^{\circ}$,c=7.185$A^{\circ}$,
        $\beta$=105.18$^{\circ}$ and in the wavelength range of 0.5 to 2$\mu$m is transparent. The most
        interesting property of NPP crystal
        is the proximity of the mean plane of molecule with the crystallographic
        plane (101); the angle between both of these planes being 11$^{\circ}$.
        Nitro group of one molecule in downward connects to Prolinol group in upper
        by hydrogen bonding. The angle between b orientation of crystal and
        N(1)-N(2) axis (charge-transfer axis) is equal to 58.6$^{\circ}$ \cite{17}. Fig.\ref{Fig2} shows the
        molecule and crystal packing of the NPP.
        For accurate and valid simulation, these properties and angles have
        to be exerted. For benzene molecule, benzene ring is a circle (see fig.\ref{Fig3});
        In NPP molecule, nitro group acts as an acceptor and the other main groups on the other side of benzene
        ring acts as a week donor (see Fig.\ref{Fig2}). Fig.\ref{Fig4} demonstrates
        the bond lengths and angles of the NPP molecule. As we see in this figure the bond lengths
        in benzene ring are not same. In our simulation, for similarity
        we consider an ellipse correspond to circle for electron cloud.
        We obtained $\varepsilon$=0.26 for the ellipse of NPP from simulation.

    \subsection{\normalsize Our Model for Electro-Optic Effect}
        Assume that single crystal
        film lies in x-y plane and light propagation is in the z
        direction; therefore in
        the presence of an electric field the equation of index
        ellipsoid by assuming crystal symmetry  will become:
        \begin{equation}
            \label{eq2}
            [(\frac{1}{n^{2}})_{_{x}}+r_{12}E_{y}]x^{2}+
            [(\frac{1}{n^{2}})_{_{y}}+r_{22}E_{y}]y^{2}+
            2r_{61}E_{x}xy=1
        \end{equation}
        that $E_{x}$ and $E_{y}$ are transverse electric field components,
        \cite{19}.
        With appropriate rotational transformation, these
        relations can been simplified. In NPP, $r_{12}$ and $r_{22}$ are large coefficients. Therefore
        $n_{x}$ and $n_{y}$ simplified to
        $n_{x}-(1/2)n_{x}^{3}r_{12}E_{y}$ and
        $n_{y}-(1/2)n_{y}^{3}r_{22}E_{y}$ respectively. The
        phase retardation $\Gamma$, with an applied electric field
        in a typical linear transverse electro-optic modulator will
        be obtained as follows, \cite{38}:
        \begin{equation}
            \label{eq4}
            \Gamma=\acute{\phi_{y}}-\acute{\phi_{x}}=\frac{\omega l}{c}\cdot[n_{y}-n_{x}-\frac{1}{2}(n_{y}^{3}r_{22}-n_{x}^{3}r_{12})E_{y}]
        \end{equation}
        The total phase difference between two
        perpendicular polarization of light (in our example $E_{x}$ and
        $E_{y}$),is
        \begin{equation}
            \label{eq6}
            \Delta=\Delta_{0}+\delta
        \end{equation}
        where $\Delta_{0}$ is due to linear birefringence and
        $\delta$ is due to linear electro-optic effect. In this
        case $\delta$ is much smaller than $\Delta_{0}$.\\
        In sub-micron space scales and sub-femtoseconds time scales,
        the optical constants loses its stabilization and classical
        equations for linear and nonlinear optical phenomena are not useful
        \cite{36}.\\
        Now, we suggest a microscopic model for linear electro-optic phenomenon.
        That would be assumed only
        linear optic phenomenon exist and
        nonlinear optic phenomenon do not exist, approximately. Because laser watt is not much,
        one photon interacts with any NPP molecules.
        In each interaction between photon and electron
        in every layer of crystal,
        we suppose a delay time equal to $\tau_{i}$ ($i$th layer of the crystal).
        Total delay time for m layers in crystal region is equal to:
        \begin{displaymath}
            \sum_{i=1}^{m}\tau_{i}
        \end{displaymath}
        Consequently required time for photon transmission in L length
        of crystal is equal to $\tau$, achieved from relation:
        \begin{eqnarray}
            \label{eq8}
            \tau=\frac{L}{\frac{c_{0}}{n}}=\frac{nL}{c_{0}}=\frac{L}{c_{0}}+\sum_{i=1}^{m}\tau_{i}
        \end{eqnarray}
        where $c_{\scriptscriptstyle{0}}$ is velocity of light in vacuum.
        By using this relation, we can relate
        macroscopic quantity $n$ to microscopic quantity $\tau_{i}$.
        In biaxial crystals, $\tau$ and consequently $n$ depends on polarization direction
        of incident light. Because dipole-field interaction conclusion
        is different for any direction of molecule. If $\tau_{x}$
        would be a microscopic delay for interaction of $x$-polarization field with dipole (or charge transfer
        action) and $\tau_{y}$ would be a microscopic delay for interaction of $y$-polarization field with dipole then the
        final phase difference between these two fields (named phase
        retardation) will be:
        \begin{equation}
            \label{eq9}
            \Delta\phi=\omega\cdot(\tau_{x}-\tau_{y})=\omega\cdot\sum_{i}(\tau_{x_{i}}-\tau_{y_{i}})
        \end{equation}
        Of course this relation give $\Delta_{0}$ of (6). We justify
        $\delta$ from our model in later subsections.

        \subsection{\normalsize Photon-Electron Interaction in Attosecond Regime}
\begin{figure}
            \centering
            \includegraphics[scale=0.9]{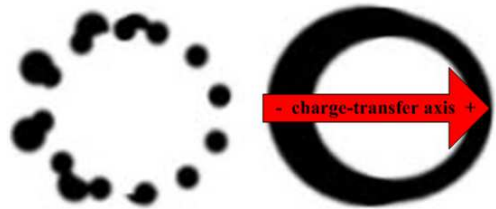}
            \caption{\footnotesize Assumed $\pi$-electron orbit of NPP molecule that obtained from Fig.\ref{Fig3} for Benzene molecule
            (approximately).}\label{Fig6}
        \end{figure}

\begin{figure}
            \centering
            \includegraphics[scale=0.7]{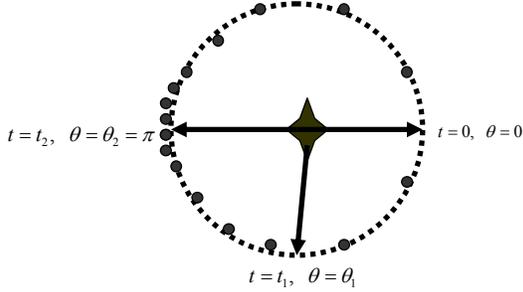}
            \caption{\footnotesize $\pi$-electron system approximation for NPP molecule, $\theta$ = 0 is in the positive
            direction of dipole (Prolinol side) and $\theta$ = $\pi$ is in the negative
            direction of dipole (Nitro side).}\label{Fig7}
        \end{figure}

\begin{figure}
            \centering
            \includegraphics[scale=0.31,width=3.1in,height=2.2in]{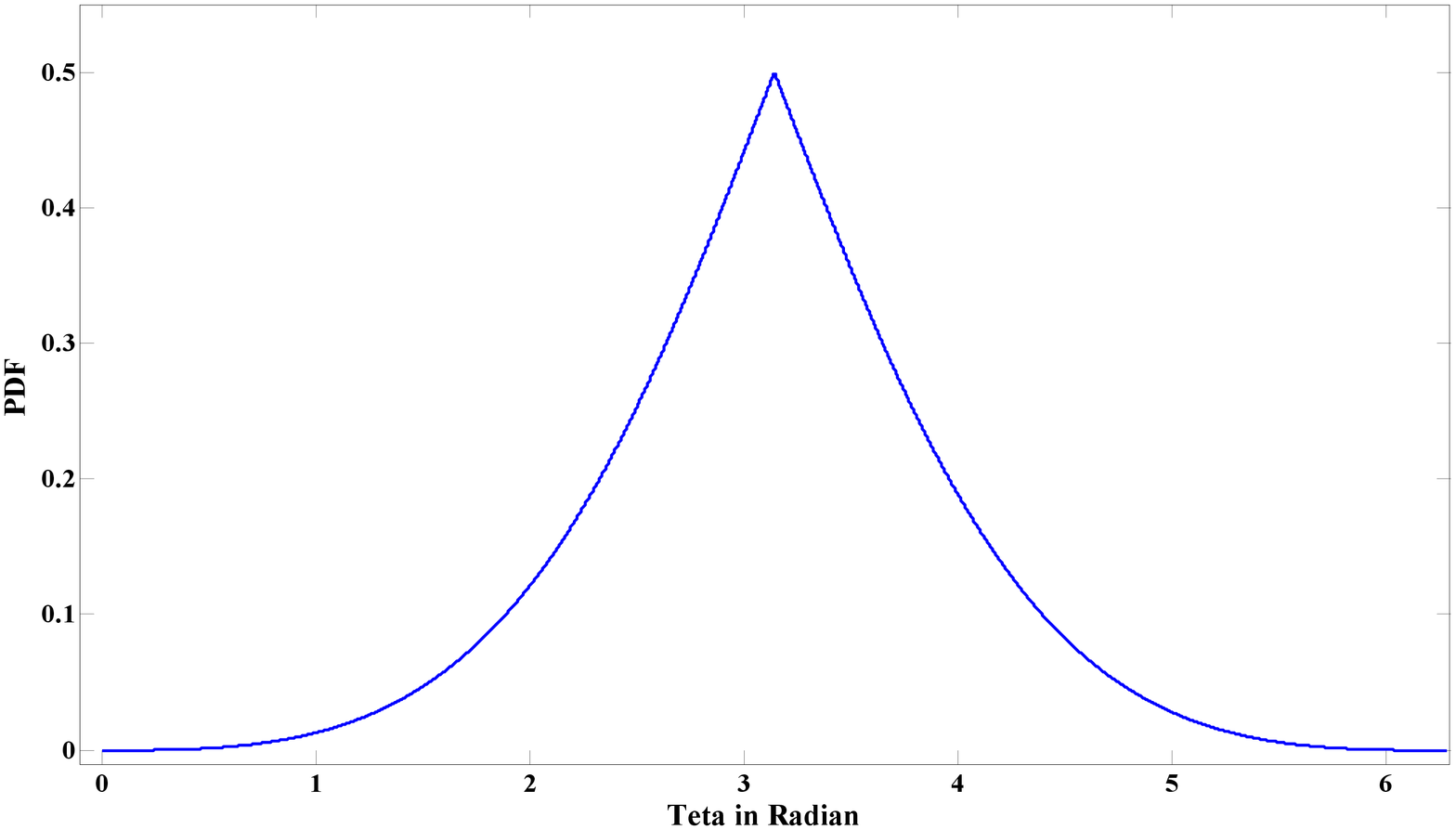}
            \caption{\footnotesize PDF of presence probability of $\pi$-electron in different energy states when it rotates around the
            assumed orbit in Fig.\ref{Fig7}; as we see the PDF is correlated to $\theta$. }\label{Fig8}
        \end{figure}

\begin{figure*}
            \centering
            \includegraphics[scale=0.5]{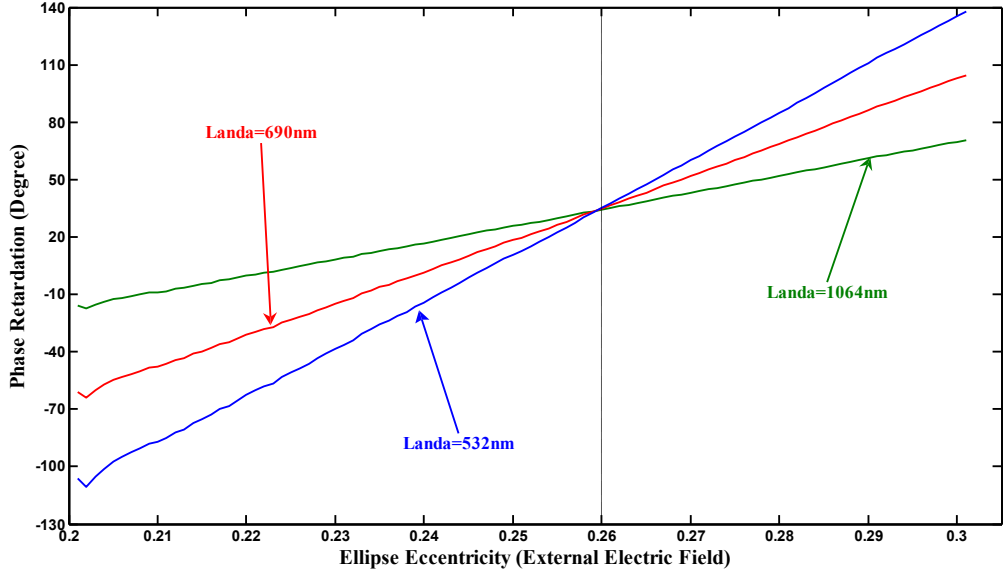}
            \caption{\footnotesize Phase retardation between $E_{x}$ and $E_{y}$ of optical signal
            with different wavelength by applied external electric field. As we
            see; by applied electric field, the virtual ellipse eccentricity change (electron in electrical field). With eccentricity
            variation, the microscopic delay ($\tau_{x}$,$\tau_{y}$) change and consequently phase retardation is modified.}\label{Fig10}
        \end{figure*}

        It can be shown that for absorbtion or emission of photons
        the material has to perform a transition between two
        eigenstates $E_{m}$ and $E_{p}$ of the material and thus
        the photon energy $E_{photon}$ has to fulfill the
        resonance condition:
        \begin{equation}
            \label{eq29}
            E_{photon}=h\nu_{photon}=|E_{p}-E_{m}|
        \end{equation}
        But for our linear phenomenon, the photon energy is about
        $2eV$ (in $\lambda=630nm$). If electron would be in HOMO (Highest Occupied
        Molecular Orbital), this electron do not go to LUMO (Lowest
        Unoccupied Molecular Orbital) or exited state by interaction.
        This phenomena is named nonresonant phenomenon,\cite{18,40,41} (nonresonant phenomena is
        not exclusive for nonlinear optical phenomena). Therefor
        electron after interaction, go to quasi states that their life times
        is very short, then  this electron go back to primary state after very short
        time. The nonresonant lifetime is determined by the
        uncertainty principle and the energy mismatch between
        photon energy in second time and the input photon energy.
        We can assume that the characteristic response time of
        this process is the time required for the electron cloud
        to  become distorted in response to an applied optical
        field. This response time can be estimated as the orbital
        period of the electron in its motion around the nucleus which is about $\tau\simeq10^{-16}s$ or
        100as,\cite{41}.
        We can estimate this characteristic response time according to (4) if, $n=3, L=3\mu$. Consequently
        $\Sigma\tau$ is equal to $10^{-14}$sec. Because in b direction of crystal in $3\mu$m length, approximately 4024
        molecules exist, therefore the average quantity of $\tau$:
        \begin{displaymath}
              \overline{\tau}=\frac{\sum\tau}{N}
        \end{displaymath}
        is in order of $10^{-18}s$ or 1 as. The perturbation in this very short time can assume semiclassically.
        In linear phenomenon in this short time, just one photon
        interacts with one molecule. Because NPP molecule has delocalization electrons,(or $\pi$-electron system),
        in benzene ring, that photon interacts with this electron type,\cite{42} and it is annihilated \cite{38}. We call this photon,
        a successful photon, (that does not produce phonon).\\
        To obtain $\pi$-electron wavefunction for benzene molecule the Schr\"{o}dinger equation may be solved.
        Since this is very complicated process, it cannot be done exactly, an approximated procedure known as H\"{u}ckle method
        must be employed. In this method, by using H\"{u}ckle Molecular-Orbital (HMO) calculation, a wave function
        is formulated that is a linear combination of the atomic orbitals (LCAO) that have overlapped \cite{37} (see
        Fig.3):
        \begin{equation}
            \label{eq31}
            \Psi=\sum_{i}C_{i}\Phi_{i}
        \end{equation}
        where the $\Phi_{i}$ refers to atomic orbitals of carbon atoms in the ring and the summation is over the six C
        atoms. The $|C_{i}|^{2}$ is the probability of the $\pi$-electron at $i$th
        atom. Thus:
        \begin{displaymath}
        \label{eq11}
            |C_{1}|^{2}+|C_{2}|^{2}+|C_{3}|^{2}+|C_{4}|^{2}+|C_{5}|^{2}+|C_{6}|^{2}=1
        \end{displaymath}
        In the case of Benzene molecule:
        \begin{displaymath}
          |C_{i}|^{2} =\frac{1}{6}
        \end{displaymath}
        as followed from the symmetry of the ring \cite{44,45}.
        But NPP molecule isn't such as Benzene molecule. NPP is polar molecule. Nitro $(NO_{2})$ is more
        powerful electronegative compound than prolinol and pulls $\pi$-electron system; consequently, the probability
        of finding $\pi$-electron system at various carbon atoms of main ring isn't the same and the probability of finding
        $\pi$-electrons near the Nitro group is greater than near the prolinol group. Therefore there is no symmetry
        for NPP and electron cloud is spindly or oblong, (similar to dom-bell) (Fig.\ref{Fig6}).
        We estimate this form of electron cloud by an ellipse
        that our calculations would be uncomplicated. We assume effective positive
        charge that is located in one of focal points of ellipse.
        The quantity of this effective positive charge is
        determined by semiclassical arguments.
        For attaining probability of electron presence on an orbit (Fig.\ref{Fig7}),
        we say, T time is required by radial vector to sweep total $\pi.u.v$
        interior area of ellipse (u and v are semimajor and semiminor axis of ellipse respectively),
        in t times, this radial vector sweeps:
        \begin{displaymath}
          \pi.u.v.\frac{t}{T}
        \end{displaymath}
        area of ellipse, (see Fig.\ref{Fig7}). If t is the time, that electron sweeps $\theta$ radian of orbit
        then t is obtained from this relation \cite{33}:
        \begin{equation}
            \label{eq39}
            t=\frac{T}{2\pi}\{2\arctan(\sqrt{\frac{1-\varepsilon}{1+\varepsilon}}\tan(\frac{\theta}{2}))-
            \frac{\varepsilon.\sqrt{1-\varepsilon^{2}}.\sin(\theta)}{1+\varepsilon.\cos(\theta)}\}
        \end{equation}
        Where $\varepsilon$ is ellipse eccentricity. By using this relation, we attain the required
        time (t) for electron to traverse from $\theta$ to $\theta+d\theta$ and it is divided
        by total time T. By this approach, we can determine the PDF (Probability Density Function) approximately. The PDF in apogee
        (near the Nitro group), is maximum and in perigee (near the Prolinol or Methane) is
        minimum. Therefore PDF is correlated to $\theta$ from (8) and seen in Fig.\ref{Fig8}.
        The $\tau$ quantity is correlated to $\theta$ and consequently, $\tau$
        quantity is correlated to presence probability of
        $\pi$-electron system.\\
        The angle between Y vector and charge transfer action (N$_{1}$-N$_{2}$) is 58.6$^\circ$ and X and Z
        axis is perpendicular to Y (Fig.\ref{Fig4}).
        We consider propagation along Z direction. We spot a photon interacts
        with $\pi$-electron of NPP in first layer. After interaction, this photon gives
        its energy to electron and is annihilated. Electron absorbs energy and digresses in direction of photon momentum.
        Electron with photon energy, may not be unbounded and after arriving to apogee of digression, it returns back to
        ground state, because the photon energy is equal to $h\nu=1.96ev$ (h is Planck's constant and $\nu$ is frequency of
        laser beam) whereas energy for excitation is greater than $3ev$. When electron returns to
        ground state one photon is produced. The time coming up and down is $\tau$ delay time. This photon after freedom goes to second
        layer in direction of annihilated photon (nonce, we assume the polarization doesn't change), in second layer
        this photon interacts with another delocalization $\pi$-electron
        certainly, because the effective interaction range of photon is approximately equal to its wavelength and is very greater than
        the distance between molecules. This molecule is nearest to photon effective central.
        This action is repeated for each layer. The location of photon-electron
        interaction is significant in every molecule and it is effective on $\tau$ quantity directly.
        We assume that interacting photon has circle polarization and electron subject to virtual positive charge
        center. The phase retardation between $E_{x}$ and $E_{y}$ (2)
        can be obtain from this relation,\cite{33}:
        \begin{equation}
            \label{eq41}
            \Delta\phi=\frac{\omega.\sqrt{2h\nu.m}}{KZe^{2}}\sum_{i=1}^{m}[\cos(\theta_{i})-\sin(\theta_{i})].r_{i}^{2}
        \end{equation}
        that
        \begin{displaymath}
            r_{i}=\frac{(1-\epsilon^{2}).u}{1+\epsilon.\cos(\theta_{i})}
        \end{displaymath}
        where $\epsilon$ is the elliptical eccentricity
        and u is the semimajor axis of the ellipse. By applying an
        external transverse electric field to organic crystal (in the range of several volts per
        micron) the shape of $\pi$-electron system will be
        deformed slightly and we would be expect some noticeable
        variations in microscopic delay parameters ($\tau_{x}$,$\tau_{y}$)
        and phase retardation; (see Fig.\ref{Fig10}).
        Consequently, by step-like change of input
        voltage, optical signal is switched between output port of
        DOS.

        \subsection{\normalsize Variation Analysis of Phase Retardation versus Applied Electric Field}
        We simulate phase retardation of $3\mu m$-length NPP crystal by Monte-Carlo method, then we generate random
        number using $MATLAB$ program. This program produces PDF quantities was explained in before subsection
        and relates each of them to every molecule. These values are indexing $\pi$-electron positions in each
        layer, by assumption a reference point (see Fig.\ref{Fig7}). Additionally we have used a $MATLAB$ program for Monte-Carlo
        simulation. The inputs of this program are:\\ \indent
        1.  The wavelength of incident optical beam in which we want to design DOSs;\\ \indent
        2.  $h, m, q, k=\frac{1}{4\pi.\varepsilon}, c_{\scriptscriptstyle{0}}$ that are Planck's constant,
        electron rest mass, elementary charge, Coulomb constant and speed of light respectively.\\ \indent
        3.  Unit cell parameters of NPP crystal: a, b, c, $\beta$ and its other parameters that have given in
        subsection (A).\\ \indent
        4.  L: crystal thickness that in our simulation it is 3$\mu$m.\\
        And the outputs of
        $MATLAB$ program are: phase retardation in each wavelength.\\ \indent
        System calibration is done semiclassically by experimental refractive index
        data.t
        In this method that we obtain three refractive indexes with x-polarization in
        threea
        with $\epsilon$ (eccentricity), u (semimajor axis of ellipse) and Z (equivalent positive charge) in a
        way that refractive indexes in three wavelengths are very close to experimental data.
        Then we would see that refractive index in other
        wavelengths and other polarization with same $\epsilon$, u and Z will be achieved. Of course these values, $\epsilon$, u and
        Z would be close to experimental structure of crystal, for example u would be greater than and smaller than
        minimum and maximum sizes of six lengths of benzene hexagonal respectively, or $\epsilon$ would be small
        but greater than zero. In other hand these values must be logical. From this method in our simulation we have
        obtained $\epsilon=0.26, Z=3.9, u=1.4A^{0}$ that is very close to experimental and structural
        data.\\ \indent
        Xu and co-workers \cite{19,26}, have done some electro-optic experiment about single crystal film
        of NPP. They have obtained
        $|n_{x}^{3}r_{12}-n_{y}^{3}r_{22}|=340 pm/V$ and
        $r_{12}=65 pm/V$ in an optical beam with 1064nm wavelength. They have studied phase retardation
        between $E_{x}$ and $E_{y}$ of optical beam as a function
        of angle between electric field and charge transfer action
        of NPP. They have concluded that the maximum phase retardation was observed for the field oriented
        along the charge-transfer axis which was parallel to the film surface. The electro-optic effect
        or phase retardation was negligible when the electric field was
        applied perpendicular to the charge-transfer axis.
        This concept could be justified by our model in previous
        subsection. When the angle between charge-transfer action
        and external electric field change, the ellipse
        eccentricity modify and consequently, the phase
        retardation alter. Obviously, from Fig.\ref{Fig6} when the external electric
        is parallel to charge-transfer axis, the ellipse drag more
        and the ellipse convert to a line. Therefor, ellipse
        eccentricity arise and from Fig.\ref{Fig10} the phase retardation
        growth. In the other hand, from Fig.5 when the angle
        between external electric field and charge-transfer axis
        change, the ellipse is gathered and convert to circle and the eccentricity
        decrease to zero. Thus from Fig.\ref{Fig10} the phase retardation is
        lowered.

        \section{\large Design}
\begin{figure}
            \centering
            \includegraphics[scale=0.6]{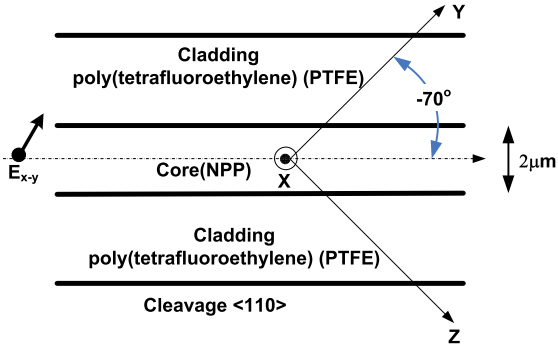}
            \caption{\footnotesize The typical waveguide that is used for our switch.
            The core of this waveguide is NPP and the cladding of
            it, is the poly(tetrafluoroethylene) (PTFE)\cite{27}. We
            assume input light that has xy-polarization and travel in
            zy-plane.}\label{Fig11}
        \end{figure}

\begin{figure}
            \centering
            \includegraphics[scale=0.5]{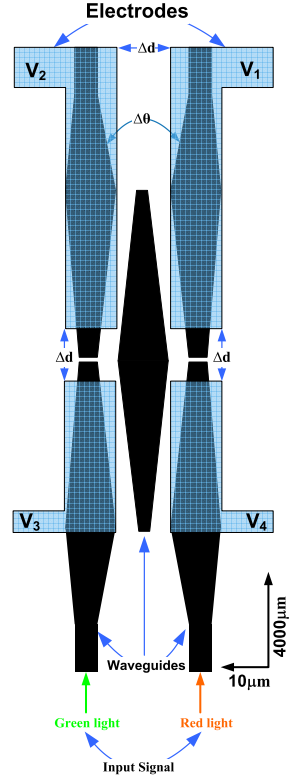}
            \caption{\footnotesize The layout of our $2\times2$ DOS with a full
            adiabatic coupler. Driving electrodes are on the top of the waveguides
            in the interaction region. By applying voltage on
            the electrodes, we can change the refractive index in one or both
            waveguides.}\label{Fig12}
        \end{figure}

\begin{figure}

            \centering
            \includegraphics[scale=0.38,width=3in,height=1.5in]{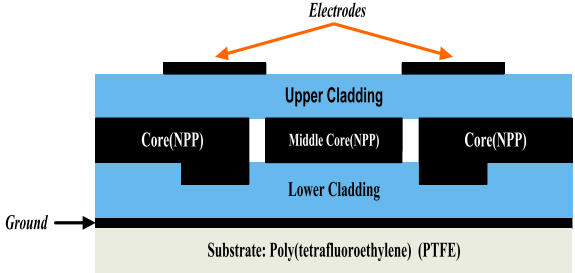}
            \caption{\footnotesize Cross section of the interaction region of our proposed DOS device. The
            driving electric field should be in the vertical direction, which
            is from the top driving electrode to the bottom electrode, to have a
            larger EO effect.}\label{Fig13}
        \end{figure}

        In this section, we design an organic $2\times2$ DOS that
        is based on EO phenomenon with adiabatic coupler. The adiabatic coupler has some advantages
        such as: a more evenly split power can be achieved in
        the switch and comparably longer electrodes can be used for the
        same device length, which leads to either lower driving voltage or
        more compact device design,\cite{3}. The basic theory of this switch
        was explained in previous section. In this switch we use a
        waveguide that has been shown in Fig.\ref{Fig11}.
        The core of this waveguide is NPP and the cladding of it,
        is the poly(tetrafluoroethylene) (PTFE),\cite{27}. We
        assume input light that has xy-polarization and travel in
        zy-plane according to Fig.\ref{Fig11}.\\ \indent
        In Fig.\ref{Fig12},
        the layout of our $2\times2$ DOS with a full
        adiabatic coupler has been shown. The input and output
        waveguides do not have been shown. Driving electrodes are on the top of the waveguides
        in the interaction region. By applying voltage on
        the electrodes, we can change the refractive index in one or both
        waveguides. If the refractive index difference between the two arms in one side
        is large enough, the power will output from the waveguide with higher refractive index.
        For these EO devices, the material is poled in the vertical direction (see Fig.\ref{Fig13}).
        Consequently, the
        driving electric field should be in the vertical direction, which
        is from the top driving electrode to the bottom electrode, to have a
        larger EO effect. To ensure the electric field has a larger
        component in the vertical direction than the horizontal, the
        spacing between the electrodes should be no less than the thickness of the device.
        The spacings in our devices are about $10\mu$m.
        Because the NPP material has large linear electro-optic
        effect, ($r_{12}=65 pm/V$, that is two times of linear electro-optic effect of nonorganic
        material) by smaller change of applied voltage, the optical
        power can switch to the neighbor waveguide, or we can
        shorten the electrode length\cite{4}. Therefore we would
        have expected that our switch is smaller than a switch
        with nonorganic material. The full angle between the two neighbor
        waveguides, $\Delta\theta$ is about 1mrad.
        From Fig.\ref{Fig10}, we see that the phase retardation
        change for different wavelengths, with same applied electric
        field. Therefore we can switch two optical beam with
        different wavelengths to two output waveguides. By suitable
        choice of step-like voltage(using an analogue to digital converter(ADC) and choosing
        two of four electrodes), we can switch specific beam
        with optional wavelength to specific output channel.\\ \indent
        The cross section of the EO DOS in the interaction region is shown in Fig.\ref{Fig13}.
        As we see, the rib waveguide is used for this switch.

        \section{\large Conclusion}
        We analyzed and designed a novel $2\times2$ DOS that is
        based on transverse EO effect. In analysis we
        justified linear EO phenomenon by QPM. This suggested physical model could be a
        powerful tool for analyzing and explaining processes that happen in waveguides with microscopic and nanoscopic sizes.
        We showed how the phase retardation between different arguments of an optical field with distinctive wavelengths can take place.
        In design, we configured a DOS with full adiabatic coupler.
        We used a ridge waveguide that has been constructed from
        NPP crystal as core. By using an ADC and applying four proper
        voltages on four electrodes, one could switch an input optical beam
        with specific wavelength to certain output channel.


\begin{thebibliography}{45}
            \small
            \bibitem{1}  Lee M H, Min Y H, Ju J J, Do J Y and Park S K 2001 {\it IEEE J. Sel. Top. Quantum Electron.} {\bf 7} 5

            \bibitem{2}  Oh M, Zhang H, Erlig H, Chang Y, Tsap B, Chang D, Szep A, Steier W H, Fetterman H R and Dalton L R 2001 {\it IEEE J. Sel.
            Top. Quantum Electron.} {\bf 7} 5

            \bibitem{3}  Yuan W, Kim S, Steier W H and Fetterman H R 2005 {\it IEEE Photon. Technol. Lett.} {\bf 17} 12

            \bibitem{4}  Yuan W, Kim S, Sadowy G, Zhang C, Wang C, Steier W H and Fetterman H R 2004 {\it Electron. Lett.}{\bf 40} 3

            \bibitem{5}  Lee S S, Shin S Y 1999 {\it Electron. Lett.} {\bf 35} 15

            \bibitem{6} Lee S S, Shin S Y 1997 {\it Electron. Lett.} {\bf 33} 4

            \bibitem{7} Ahn J T, Park S, Do J Y, Lee J M, Lee M H and Kim K H 2004 {\it IEEE Photon.Technol. Lett.} {\bf 16} 6

            \bibitem{8} Silberberg Y, Permutter P and Baran J 1987 {\it Appl. Phys. Lett.} {\bf 51}

            \bibitem{9} Feehrer J R 1996 {\it J. Lightwave Technol.} {\bf 14} 12

            \bibitem{10} Zuo Y, Bahamin B, Tremblay E J, Pulikkaseril C, Shoukry E, Mony M, Langlois P, Aimez V and Plant
            D V 2005 {\it IEEE Photon. Technol. Lett.} {\bf 17} 10

            \bibitem{11} Li J, Myrén N, Margulis W, Ortega B, Puerto G, Pastor D, Capmany J, Belmonte M
            and Pruneri V 2005 {\it IEEE Photon. Technol. Lett.} {\bf 17} 12

            \bibitem{12} Glebov A L, Lee M G, Huang L, Aoki S, Yokouchi K, Ishii M and Kato M 2005 {\it IEEE J. Sel. Top. Quantum Electron.} {\bf 11} 2

            \bibitem{13} Massaro A, Cavallari F and Rozzi T 2004 {\it J. Lightwave Technol.} {\bf 22} 2

            \bibitem{14} Ellis A D, Widdowson T, Shan X and Moodie D G 1994 {\it Electron. Lett.} {\bf 30} 16

            \bibitem{15} Song X, Futakuchi N, Yit F C, Zhang Z and Nakano Y 2005 {\it IEEE Photon. Technol. Lett.} {\bf 17} 7

            \bibitem{16} Takada A and Park J H 2002 {\it J. Lightwave Technol.} {\bf 20} 12

            \bibitem{17} Zyss J,  Nicoud J F and Coquillay M 1984 {\it J. Chem. Phys.} {\bf 81} 4160

            \bibitem{18} Nalwa H and Miyata 1997 {\it Nonlinear optics of organic molecules and polymers} (Hitachi research
              Laboratory and Tokyo University of Agriculture and Technology)

            \bibitem{19} Xu J, Zhou L and  Thakur M 1996 {\it Appl. Phys. Lett.} {\bf 69} 1197

            \bibitem{20} Ledoux I, Lepers C, Perigaud A, Badan J and Zyss J 1990 {\it Opt. Commun.} {\bf 80} 149

            \bibitem{21} Ledoux I, Josse D, Vidakovic P and Zyss J 1986 {\it Opt. Eng.} {\bf 25} 202

            \bibitem{22} Banfi G P, Datta P K, Degiorgio V, Fortusini D, Shepherd E E A and Sherwood J N 1999 {\it Chem. Phys.} {\bf 245}

            \bibitem{23} Datta P K, Fortusini D, Donelli G, Banfi G P, Degiorgio V, Sherwood J N and Bhar G C 1998 {\it Opt. Commun.} {\bf 149}

            \bibitem{24} Banfi G P, Degiorgio V, Sherwood J N 2001 {\it Synthetic Metals} {\bf 124}

            \bibitem{25} Quintero-Torres R and Thakur M 1996 {\it Appl. Phys. Lett.} {\bf 69}

            \bibitem{26} Xu J, Zhou L and  Thakur M 1998 {\it Appl. Phys. Lett.} {\bf 72}

            \bibitem{27} Vallee R, Damman P, Dosiere M and Zyss J 2001 {\it J. Chem. Phys.} {\bf 115}

            \bibitem{28} Levine B F, Bethea C G, Thurmond C D, Lynch R T and Bernstein J L 1979 {\it J. Appl. Phys.} {\bf 50} 4

            \bibitem{29} Khanarian G, Che T, Demartino R N, Haas D, Leslie T, Man H T and Sasone M 1987
            {\it SPIE Advances in Nonlinear Polymers and Inorganic Crystals, Liquid Crystals and Laser Media} {\bf 824}

            \bibitem{30} Ho E S S, Iizuko K, Freundorfer A P and Wah C K L 1991 {\it J. Lightwave Technol.} {\bf 9} 1

            \bibitem{31} Williams D J Ed. 1983 {\it Nonlinear Optical Properties of Organic and
            Polymeric Materials} {American Chemical Society, ACS symposium series}

            \bibitem{32} Ohtsu M, Kobayashi K, Kawazoe T, Sangu S and Yatsui T 2002 {\it IEEE J. Sel. Top. Quantum Electron.} {\bf 8} 4

            \bibitem{33} Kaatuzian H and Wahedy Zarch A A 2004 {\it Proc. of CSIMTA Cherbourg-Normandy-FRANCE}

            \bibitem{34} Kaatuzian H, Bazhdanzadeh N and Ghohrodi Ghamsari B 2004 {\it Proc. of CSIMTA Cherbourg-Normandy-FRANCE}

            \bibitem{35} Kaatuzian H and Wahedy Zarch A A 2006 {\it Proc. CSNDSP2006, Patras Univ, Greece.}

            \bibitem{36} Adibi A and Kaatuzian H 1995 {\it J. of Engineering, Islamic Republic of Iran} {\bf 8} 4

            \bibitem{37} March J 1992 {\it Advanced Organic Chemistry: Reactions, Mechanisms and Structure} (4th Ed. John Wiley and Sons)

            \bibitem{38} Yariv A 1975 {Quantum Electronics(2nd Ed.)} John Wiley and Sons Inc

            \bibitem{39} Saleh B E A and Teich 1991 {\it Fundamentals of Photonics} (2nd Ed. John Wiley and Sons)

            \bibitem{40} Menzel R 2000 {\it Photonics, Linear and nonlinear interactions of laser light and matter} Springer

            \bibitem{41} Boyd R W 2003 {\it Nonlinear optics(3th Ed.)} John Wiley and Sons Inc.

            \bibitem{42} Dirk C W, Twieg R J and wagniere G 1986 {\it J. Am. Chem. Soc.} {\bf 108} 18

            \bibitem{43} Lalama S J and Garito A F 1979 {\it Phys. Rev. A} {\bf 20} 3

            \bibitem{44} Omar M A 1974 {\it Elementary Solid State Physics}(Mills and Bons Ltd.)
            \bibitem{45} Waddington N 1972 {\it Modern Organic Chemistry} (Addison-Wesley)
\end{thebibliography}
 \end{document}